 \documentclass[pmlr,twocolumn,10pt]{jmlr} 

\usepackage{booktabs}
\usepackage{siunitx}
\usepackage{multirow}

\newcommand{\equal}[1]{{\hypersetup{linkcolor=black}\thanks{#1}}}

\newcommand{\mL}{\mathcal{L}}
\newcommand{\mR}{\mathbb{R}}
\newcommand{\bx}{\mathbf{x}}
\newcommand{\by}{\mathbf{y}}
\newcommand{\bz}{\mathbf{z}}

\theorembodyfont{\upshape}
\theoremheaderfont{\scshape}
\theorempostheader{:}
\theoremsep{\newline}

\jmlrvolume{LEAVE UNSET}
\jmlryear{2023}
\jmlrsubmitted{LEAVE UNSET}
\jmlrpublished{LEAVE UNSET}
\jmlrworkshop{Machine Learning for Health (ML4H) 2023} 

\title[Multi-task learning for Base Editor Outcome Prediction]{Attention-based Multi-task Learning for Base Editor Outcome Prediction}

\author{%
\Name{Amina Mollaysa}\equal{These authors contributed equally} \Email{maolaaisha.aminanmu@uzh.ch}\\
\addr University of Zurich, Switzerland
\AND
\Name{Ahmed Allam}\footnotemark[1] \Email{ahmed.allam@uzh.ch}\\
\addr University of Zurich, Switzerland
\AND
\Name{Michael Krauthammer} \Email{michael.krauthammer@uzh.ch}\\
\addr University of Zurich, Switzerland
}

\begin{document}

\maketitle

\begin{abstract}
Human genetic diseases often arise from point mutations, emphasizing the critical need for precise genome editing techniques. Among these, base editing stands out as it allows targeted alterations at the single nucleotide level. However, its clinical application is hindered by low editing efficiency and unintended mutations, necessitating extensive trial-and-error experimentation in the laboratory.  To speed up this process, we present an attention-based two-stage machine learning model that learns to predict the likelihood of all possible editing outcomes for a given genomic target sequence.  We further propose a multi-task learning schema to jointly learn multiple base editors (i.e. variants) at once. Our model's predictions consistently demonstrated a strong correlation with the actual experimental results on multiple datasets and base editor variants. These results provide further validation for the models' capacity to enhance and accelerate the process of refining base editing designs. 
\end{abstract}
\begin{keywords}
Base editor, self-attention, multi-task learning, CRISPR, genome editing
\end{keywords}

\section{Introduction}
The landscape of human genetic diseases is largely dominated by a significant proportion of cases arising from point mutations \citep{landrum2014clinvar}. These minor genetic alterations (substitution, deletion, or insertion of a single nucleotide), pose serious implications for health ranging from rare monogenic conditions to more common chronic diseases. 
Genome editing approaches allow researchers to make targeted changes to the DNA of living cells. Among these, base editing
\citep{komor2016programmable} shows promising results as it enables precise genome editing at the single nucleotide level without causing double-stranded breaks (DSBs) in the DNA. 
While base editors have great potential as genome editing tools for basic research and gene therapy, their application has been limited due to 1) low editing efficiency on specific sequences or 2) unintended editing results with concurrent mutations, especially where there are multiple substrate nucleotides within close proximity to the intended edit.

Developing a robust machine learning model capable of accurately predicting the potential editing outcomes of diverse base editors on various target sites could significantly enhance the field. It allows the biologists to assess possible outcomes much faster and fine-tune their editing strategies with high efficiency. 
In this paper, we focus on estimating the probability of potential outcome sequences when various base editors are applied to specific DNA targets. 
Different editors exhibit varying behaviors when applied to the same target sequences due to factors such as binding affinities and editing window sizes leading to distributional shifts. Rather than training individual models for each editor, we propose a multi-task learning framework to train a unified model and learn from multiple editors simultaneously. 
We train and test our models on six datasets corresponding to the experimental outcomes from six base editors applied on thousands of target sites (Table \ref{tab:data}). 


\section{Method}
\paragraph{Base editor}
Base editors (BEs) are created by fusing the Cas9 protein with DNA-modifying enzymes. They are directed by a 20-base pair guiding RNA molecule (sgRNA) that acts as a GPS to locate and bind to a matching DNA segment known as the protospacer. The effectiveness of BEs largely depends on the composition of this protospacer sequence. BEs, in tandem with the sgRNA, can only bind to the DNA if there's a protospacer adjacent motif (PAM) - a sequence consisting of 2-6 nucleotides - present adjacent to the protospacer. This PAM sequence further influences the activity of BEs (see section \ref{apend:base_editor}). 

\begin{figure}[tb]
    
    \centering
    \includegraphics[scale=0.23]{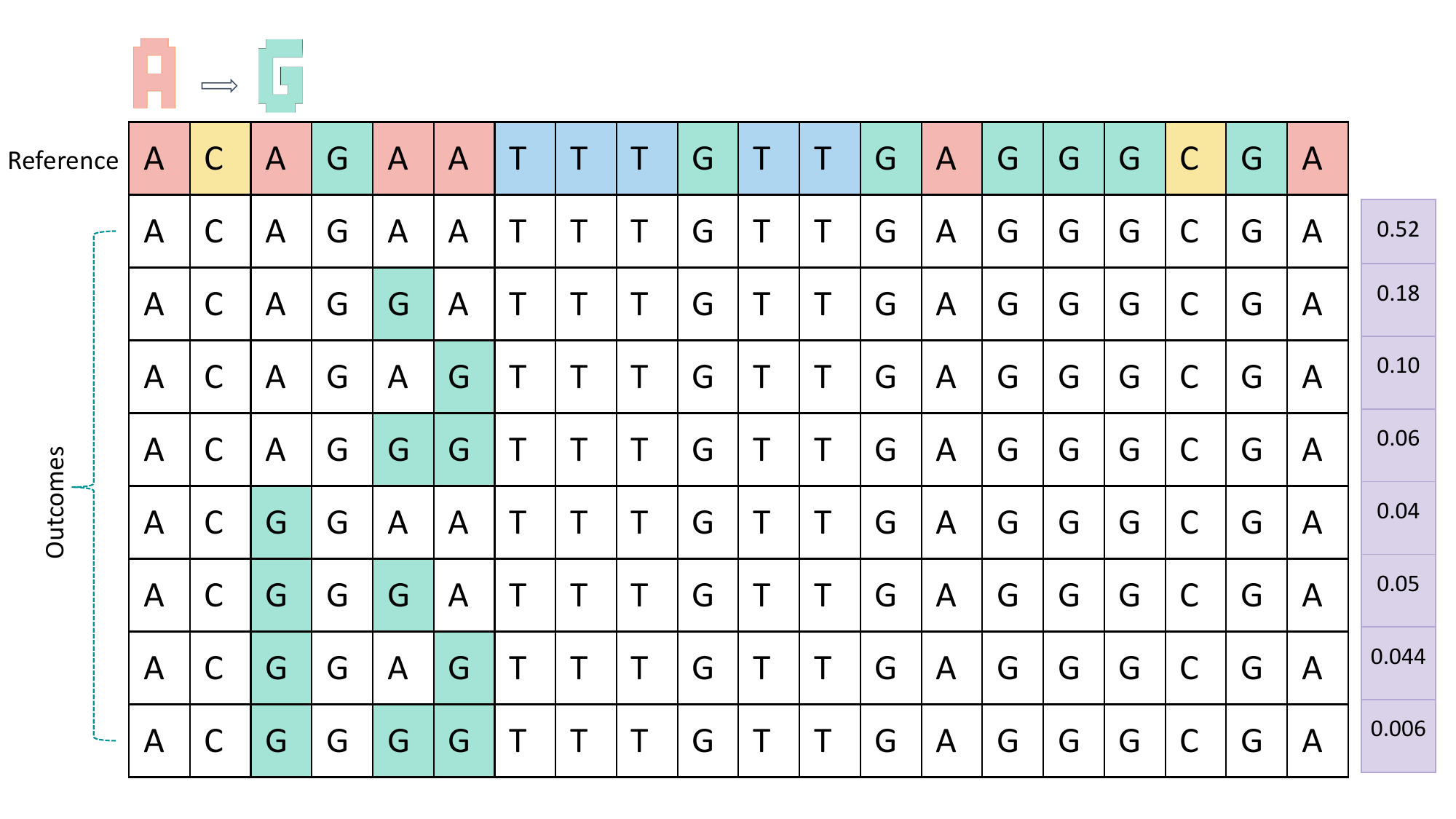}
    \caption{\small{
    An example of a reference sequence of 20 bases (i.e. nucleotides) and associated outcome sequences when applying ABEmax base editor. The first row represents the reference (target) sequence, and the second row is the outcome sequence with no modification (i.e. wild-type) with a probability of occurrence of 0.52. The third row represents a possible outcome sequence where the letter A is changed to G at position 5 with a probability of 0.35. The rest of the rows represent all possible changes of the reference sequence targeting letter A to G with their associated probabilities.}}
    \label{fig:data_demo}
\end{figure}
\begin{figure}
    \centering
    \centering
    \includegraphics[scale=0.35]{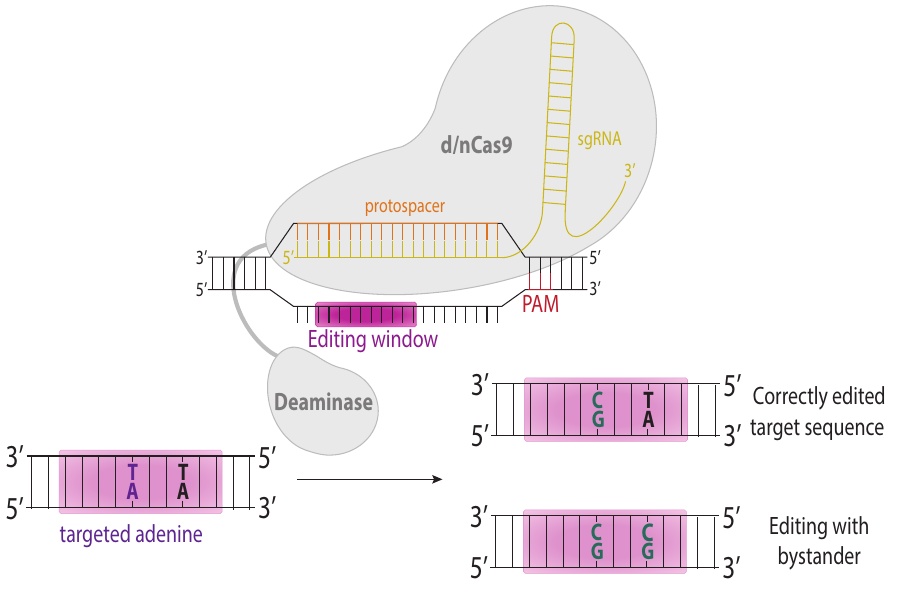}
    \caption{\small{Adenine base editor}} 
    \label{fig:base_editor}
\end{figure}

\paragraph{ Data representation} 
Assume we have a target/reference DNA sequence denoted by $\bx_{\text{ref}} = [x_1, x_2,\dots, x_T]$ where $x_i\in \{A, C, G, T\}$, and a  set of DNA sequences $ \mathbf{X}_{\text{out}} = [\bx_{\text{out},1}, \bx_{\text{out},2}, \dots, \bx_{\text{out}, M}]\in \mR^{M\times T}$ representing corresponding outcomes when a specific base editor is applied to the reference sequence $\bx_{\text{ref}}$. The associated probabilities for these outcomes are given by $\by=[y_1, y_2, \dots, y_{M}]$ where $y_i= P(\bx_{\text{out}, i}|\bx_{\text{ref}}) \in [0,1], \: \textit{for} \:\:i = 1, 2, \dots, M$,  indicating the likelihood of obtaining outcome $\bx_{\text{out},i}$ through editing of $\bx_{\text{ref}}$. Here, $T$ is the reference sequence length, and $M$ is the total possible outcomes, which vary with the reference sequence. Figure \ref{fig:data_demo} shows an example of a reference sequence and its outcomes.
To represent the reference sequence, 
we consider protospacer, PAM, and overhangs ( Figure \ref{fig:base_editor}). Here, ``overhangs" refer to adjacent nucleotides on both sides of the protospacer. 
For simplicity, we use $\bx_{\text{ref}}$ to denote the reference sequence which could refer to one of these representations: (a) protospacer, (b) protospacer + PAM, or a (c) left overhangs + protospacer + PAM + right overhangs where + is the concatenation operator. Respectively, the outcome sequences match the reference sequence in length but differ in the modified target bases in the protospacer. The outcome sequence identical to the reference sequence (with no edits) is referred to as the wild-type. The training dataset comprises $N$ pairs, each containing a reference sequence, its associated outcomes, and the corresponding probabilities, denoted by $D = \{\bx_{\text{ref}}^i, \mathbf X_{\text{out}}^i, \by^i\}_{i=1}^N$. To simplify, we omit instance-level indexing and use only $\bx_{\text{ref}}$ when referring to a specific reference sequence.

\subsection{Problem formulation}\label{seq: model_formulation}
Our objective is to predict the likelihood of potential outcomes resulting from a specific BE applied to a reference sequence. One approach would be formulating it as a generative model where we directly model the conditional distribution $P(X_{\text{out}}|\bx_{\text{ref}})$ that we can use to sample different outcomes for a given reference sequence and calculate the probability of each outcome. However, unlike typical generative models that must learn to generate entire output sequences, our scenario benefits from already knowing a portion of the output sequences.
Due to the BEs specific targeting of A-to-G or C-to-T transformations, a substantial portion of the output sequence remains consistent with the reference sequence, with only a few positions undergoing alteration.


In the inference phase, for a given reference sequence, we can efficiently generate all possible outcomes by considering only the edit combination of target bases (A/G) within the protospacer. Therefore, we only need to learn the distribution $P(X_{\text{out}}|\bx_{\text{ref}})$ such that we can evaluate the probability of a specific outcome for a given reference sequence $P(X_{\text{out}} = \bx_{\text{out}, i}|\bx_{\text{ref}})$. However, there is a relatively higher probability often associated with the wild-type outcome (not edited) 
compared to the edited outcomes. This situation presents a challenge when directly modeling $P(X_{\text{out}}|x_{\text{ref}})$— as the model might easily learn the wild-type probability but struggle with outcomes that have extremely low probabilities.

\subsection{Two-stage model}
Therefore, we propose a two-stage model where we break down $P(X_{\text{out}}|\bx_{\text{ref}})$ as the product of two probabilities:
\begin{multline}
P(\bx_{\text{out},i}|\bx_{\text{ref}}) =
\begin{cases}
    P(\bx_{\text{out}, i}|\bx_{\text{ref}}, \text{edited})P(\text{edited}|\bx_{\text{ref}}), \\ \:\:\:\:\:\:\:\:\:\:\:\:\:\:\:\:\:\:\:\:\:\:\:\:\:\:\:\:\:\:\:\:\:\:\:\:\:\text{if } \bx_{\text{out}, i} \neq \bx_{\text{ref}} \\
    1 - P(\text{edited}|\bx_{\text{ref}}), \text{if } \bx_{\text{out}, i} = \bx_{\text{ref}}
\end{cases}
\label{model:full}
\end{multline}
For a given reference sequence, we first predict the probability of \textit{overall efficiency} (Eq.\ref{eq:efficiency}). It provides the probability of the target sequence being edited, $P(edited|\bx_{\text{ref}})$. 
Next, we predict the probability of all possible edited outcomes, $P(\bx_{\text{out}, i}|\bx_{\text{ref}}, edited)$. 
\begin{equation}
\resizebox{0.9\linewidth}{!}{
 $P(edited|\bx_{\text{ref}})= \frac{\textit{Sum of the read count of all edited reads for the target}}{\textit{Total read count of the target sequence}}$
 }
 \label{eq:efficiency}
\end{equation}
We estimate the \textit{overall efficiency} of the given reference sequence using $f_{\mathbf{\theta}_1}(\bx_{\text{ref}})$, denoted by the \textit{overall efficiency} model, and the conditional probabilities of all non wild-type outcomes using $f_{\mathbf{\theta}_2}(\bx_{\text{ref}},\bx_{\text{out}, i})$  which we denote by \textit{proportion model}.

\begin{equation}
 f_{\mathbf{\theta}_1}(\bx_{\text{ref}}) = P(edited|\bx_{\text{ref}})
 \label{eq:model_overalleff}
 \end{equation}
 \begin{equation}
 f_{\mathbf{\theta}_2}(\bx_{\text{ref}},\bx_{\text{out}, i}) = P(\bx_{out, i}|\bx_{\text{ref}}, \textit{edited}), \:\: 
  \label{eq:model_bystander}
\end{equation}
$\text{where}\:\:  \bx_{\text{out}, i}\neq \bx_{\text{ref}}$. Once $f_{\mathbf{\theta}_1}$ and $f_{\mathbf{\theta}_2}$ are learned,  we can calculate $P(X=\bx_{\text{out}, i}|\bx_{\text{ref}})$ where $i = 1, \dots M$ for all outcome sequences, including wild-type and edited sequences using Eq \ref{model:full}.
\begin{figure}[tb]
    \centering
    \includegraphics[scale=0.5]{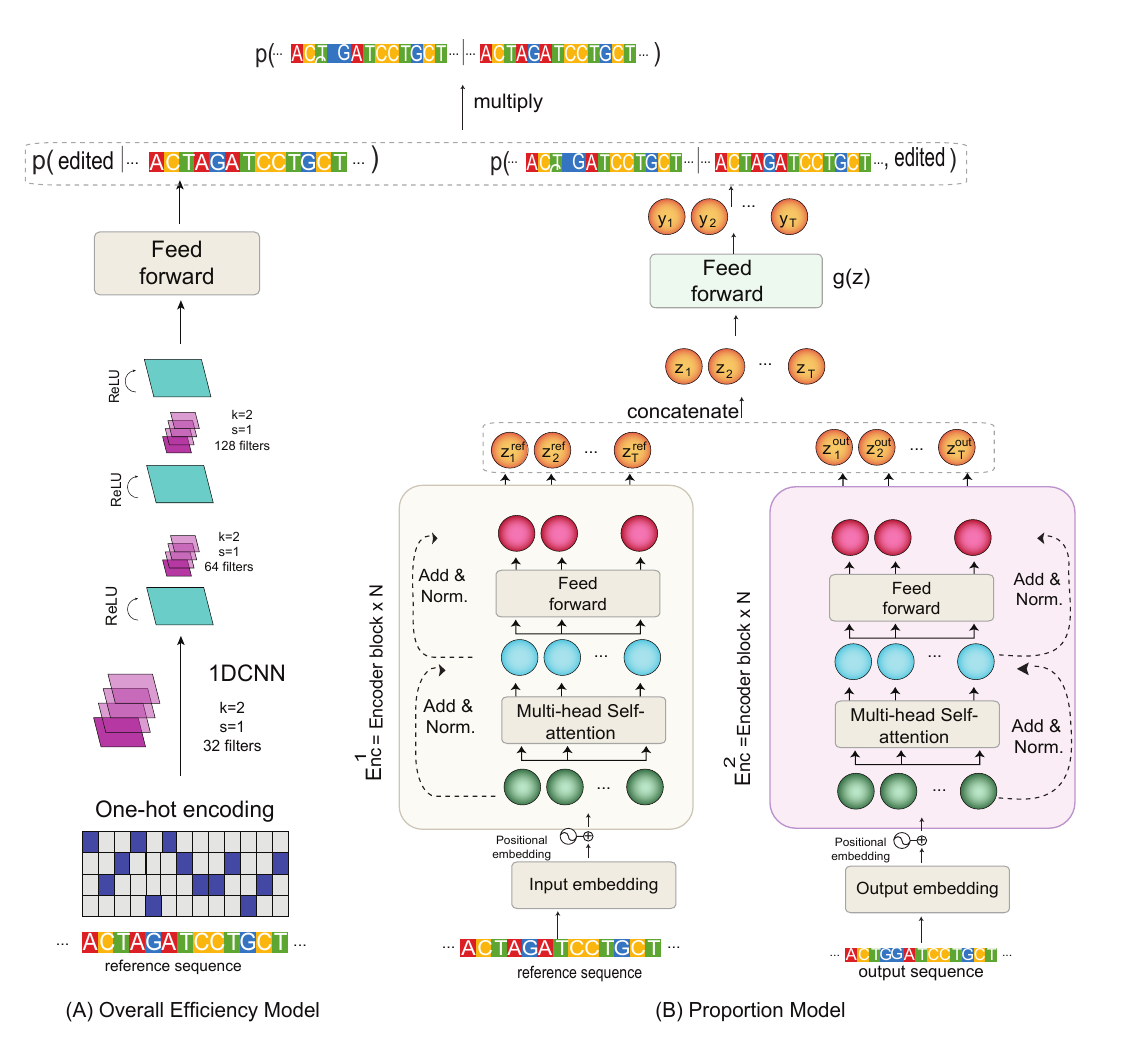}
    \caption{Two-stage Model overview}
    \label{fig:Two-step-Model}
\end{figure}
The final objective is composed of both losses (KL divergence measure) from the overall efficiency and proportion models (see appendix \ref{sec:twostage}),  with a weight regularization term on the model parameters represented by $\mathbf{\theta} = \{\mathbf{\theta_1}, \mathbf{\theta_2}\}$:
\begin{equation}
  \mL_{\textit{proportion}}(\mathbf{\theta_1}; D) + \mL_{\textit{efficiency}}(\mathbf{\theta_2}, D)+ \frac{\lambda}{2}\|\mathbf \theta\|_2^2
  \label{eq:composite_loss_func}
\end{equation}

\subsection{Multi-task learning with multiple BEs}\label{sec:Multi_task_learning}
\begin{figure}
    \centering
    \includegraphics[scale=0.5]{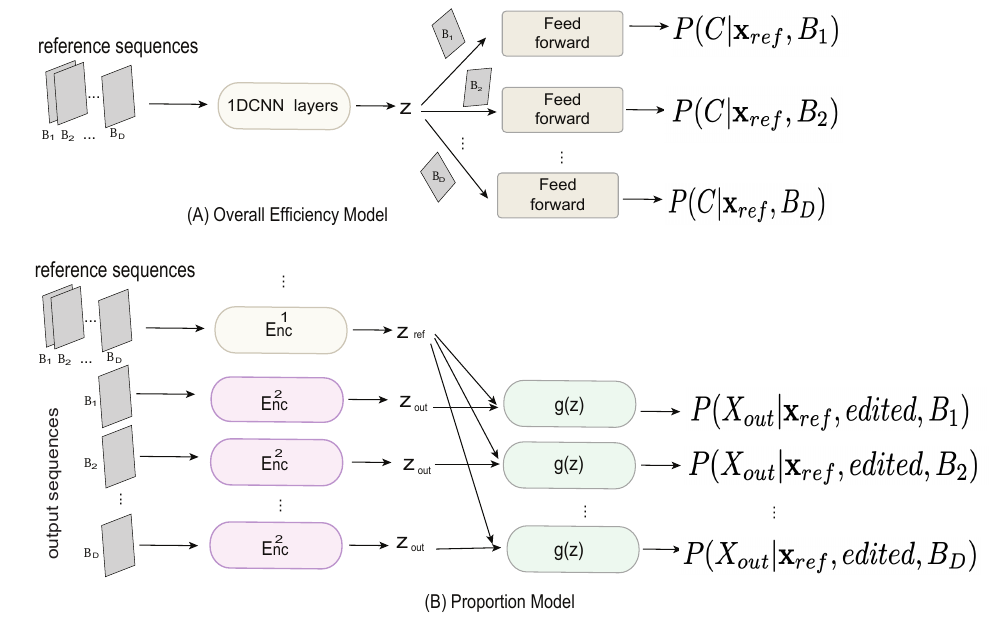}
    \caption{Multi-task learning model overview}
    \label{fig:multi_task}
\end{figure}
There exists a diverse set of BEs, each distinguished by its unique design attributes, resulting in different editing efficiencies when applied to a given target sequence.
Conventional approaches have often proposed training separate models for each editor. 
To leverage common patterns and relationships present across various datasets derived from various BEs, 
and reduce computational time, we propose a more efficient solution based on multi-task learning. Instead of training separate models for each editor, we train a single model capable of predicting the efficiency of various editors simultaneously. 

Given a total number of $D$ editors where each editor has its own dataset $B_i$, we developed a multi-task learning model that uses shared encoding layers to extract a common representation across all the datasets as well as individual branches that fine-tune the model specifically for each library, ensuring a better fit to their respective data. 
 This approach implicitly models $P(X_{\text{out}}\:|\:\bx_{\text{ref}}, B_i)$ where $B_i$ represents the base editor type applied on the reference sequence.
To implement multi-task learning across all datasets,  we extend our proposed two-stage model architecture for multi-task learning, as depicted in Figure \ref{fig:multi_task}. 


\section{Experiments}
\paragraph{Dataset} To comprehensively assess BEs efficiency across thousands of genomic sequences, we conducted high-throughput screening, resulting in the creation of six distinct datasets. Each dataset corresponds to the application of one of the following base editors: SpRY-ABE8e, SpCas9-ABE8e SpG-ABE8e, SpRY-ABEmax, SpCas9-ABEmax, and SpG-ABEmax (see appendix Table \ref{tab:data}). Detailed descriptions of the used editors are provided in Appendix Section \ref{appendix:BE}. 


\paragraph{Experiment setup}
All reported results are based on the average performance over the three runs(indicated by $mean \pm std$). We use a one-stage model (appendix \ref{sec:onestage}) that computes $P(X_{out}|\bx_{ref})$ using the proportion model architecture of the two-stag model as a baseline.  In the first step, we use this one-stage model to identify optimal features for representing the target/reference sequence. Subsequently, utilizing these selected features (i.e., protospacer + PAM, as detailed in appendix \ref{sec:input_rep}), we compare the two-stage model's performance against the one-stage model. Finally, we extend the two-stage model to multi-task learning, comparing it with single-task learning, where separate models are trained for different editors.
\subsection{Experiment results}

\paragraph{Two-stage Model} 

\begin{table}[tb]
    \centering
     \resizebox{0.8\linewidth}{!}{%
    \begin{tabular}{lllll}
    \toprule
    & \multicolumn{2}{c}{One-stage Model}& \multicolumn{2}{c}{Two-stage Model}\\
      Libraries & Spearman& Pearson&Spearman&Pearson\\
        
         \hline
         SpRY-ABEmax &$0.854\pm 0.006$& $0.983\pm 0.001$& $0.873 \pm 0.001$ & $ 0.986 \pm 0.001$ \\
        
         SpCas9-ABEmax &$0.881\pm 0.006$&$0.989\pm 0.0005$  & $0.879 \pm 0.004$ & $0.991 \pm 0.001$ \\

         SpG-ABEmax &  $0.866 \pm 0.004$ & $0.989 \pm 0.0003$& $0.887\pm 0.003$ &$0.991\pm 0.0006$\\
        SpRY-ABE8e &  $0.779 \pm 0.003$ & $0.968 \pm 0.002$ &$0.862\pm 0.003$ &$0.974\pm 0.001$\\
        SpCas9-ABE8e &$0.857\pm 0.007$&$0.945\pm 0.0006$& $0.856 \pm 0.003$ & $0.937 \pm 0.002$  \\
        SpG-ABE8e & $0.820\pm 0.005$&$0.974\pm 0.0009$ &$0.865 \pm 0.004$ & $0.978 \pm 0.0008$ \\
    \bottomrule
    \end{tabular}}
    \caption{\small{Performance comparison between One-stage and Two-stage models on all outcomes (i.e. including wild-type sequences).}}
    \label{tab:two-step-model_overlall}

    \centering
     \resizebox{0.8\linewidth}{!}{%
    \begin{tabular}{lllll}
    \toprule
    & \multicolumn{2}{c}{One-stage Model}& \multicolumn{2}{c}{Two-stage Model}\\
      Libraries & Spearman& Pearson&Spearman&Pearson\\
        
         \hline
         SpRY-ABEmax &$0.745\pm 0.015$& $0.711\pm 0.011$& $0.799\pm 0.007$ & $0.782 \pm 0.012$ \\
        
         SpCas9-ABEmax &$0.82\pm 0.0003$&$0.851\pm 0.014$  & $0.838 \pm 0.009$ & $0.890 \pm 0.030$ \\

         SpG-ABEmax &  $0.807 \pm 0.003$ & $0.752 \pm 0.014$& $0.845\pm 0.011$ &$0.822\pm 0.014$\\
        SpRY-ABE8e &  $0.393 \pm 0.021$ & $0.508 \pm 0.025$ &$0.547\pm 0.056$ &$0.669\pm 0.051$\\
        SpCas9-ABE8e &$0.855\pm 0.007$&$0.840\pm 0.003$& $0.866 \pm 0.0021$ & $0.858 \pm 0.021$  \\
        SpG-ABE8e & $0.712\pm 0.002$&$0.732\pm 0.004$ &$0.774 \pm 0.005$ & $0.810 \pm 0.009$ \\
    \bottomrule
    \end{tabular}}
    \caption{\small{Performance comparison between One-stage and Two-stage model on wild-type outcomes only}}
    \label{tab:two-step-model_wild}
    \centering
     \resizebox{0.8\linewidth}{!}{%
    \begin{tabular}{lllll}
    \toprule
    & \multicolumn{2}{c}{One-stage Model}& \multicolumn{2}{c}{Two-stage Model}\\
      Libraries & Spearman& Pearson&Spearman&Pearson\\
        
         \hline
         SpRY-ABEmax &$0.740\pm 0.007$& $0.778\pm 0.012$& $0.798\pm 0.003$ & $0.818 \pm 0.006$ \\
        
         SpCas9-ABEmax &$0.683\pm 0.0003$&$0.748\pm 0.022$  & $0.728 \pm 0.006$ & $0.795 \pm 0.006$ \\

         SpG-ABEmax &  $0.729 \pm 0.0043$ & $0.744 \pm 0.004$& $0.778\pm 0.005$ &$0.810\pm 0.004$\\
        SpRY-ABE8e &  $0.707 \pm 0.010$ & $0.816 \pm 0.006$ &$0.809\pm 0.004$ &$0.849\pm 0.003$\\
        SpCas9-ABE8e &$0.684\pm 0.007$&$0.729\pm 0.008$& $0.714 \pm 0.014$ & $0.753 \pm 0.007$  \\
        SpG-ABE8e & $0.719\pm 0.004$&$0.787\pm 0.005$ &$0.789 \pm 0.004$ & $0.826 \pm 0.003$ \\
    \bottomrule
    \end{tabular}}
    \caption{\small{Performance comparison between One-stage and Two-stage model performance on non wild-type outcomes (i.e. edited outcome sequences)}}
\label{tab:two-step-model_nonwild}
    \centering
     \resizebox{0.8\linewidth}{!}{%
    \begin{tabular}{lllll}
    \toprule
    & \multicolumn{2}{c}{Single task learning}& \multicolumn{2}{c}{Multi task learning}\\
      Libraries & Spearman& Pearson&Spearman&Pearson\\
        
         \hline
         SpRY-ABEmax & $0.877 \pm 0.001$ & $0.986 \pm 0.001$ & $0.872 \pm 0.002$ & $0.986 \pm 0.0002$ \\
        
         SpCas9-ABEmax & $0.879 \pm 0.004$ & $0.989 \pm 0.001$ & $0.864 \pm 0.0019$ & $0.992 \pm 0.0001$ \\

         SpG-ABEmax & $0.882 \pm 0.001$ & $0.991 \pm 0.0006$ & $0.889 \pm 0.0016$ & $0.992 \pm 0.0004$ \\
        SpRY-ABE8e & $0.861 \pm 0.0029$ & $0.974 \pm 0.001$ & $0.863 \pm 0.0011$ & $0.975 \pm 0.001$ \\
        SpCas9-ABE8e & $0.856 \pm 0.008$ & $0.938 \pm 0.0005$ & $0.852 \pm 0.002$ & $0.937 \pm 0.003$ \\
        SpG-ABE8e & $0.865 \pm 0.004$ & $0.980 \pm 0.0008$ & $0.871 \pm 0.003$ & $0.979 \pm 0.001$ \\
    \bottomrule
    \end{tabular}}
    \caption{\small{Multi-task model VS single task models}}
    \label{tab:muti-task}
    
\end{table}
Table \ref{tab:two-step-model_overlall} demonstrates slightly better Spearman correlation results for the two-stage model compared to the one-stage model. This improvement is attributed to the two-stage model's approach, which initially predicts the wild-type and then refines predictions for edited outcomes.  We further evaluated each model's performance separately for both wild-type and edited outcomes. The two-stage model outperforms the one-stage model in most of the datasets when evaluating the performance on wild-type and edited outcomes separately (see Table \ref{tab:two-step-model_wild} and \ref{tab:two-step-model_nonwild}).
\paragraph{Multi-task learning}
We compared the performance of multi-task learning model (see  Figure \ref{fig:multi_task}) across all the datasets/editors with a single-task setup where we trained one model per editor.  Table \ref{tab:muti-task} reports similar performance for both models. Although there wasn't a substantial performance difference, adopting a unified multi-task model offers advantages such as reduced run-time (for training and inference) and smaller model size (fewer parameters) while maintaining consistent performance across all datasets. Moreover, with a unified model, we can simultaneously predict the editing outcomes of all six editors at once for a given target sequence.


\paragraph{Comparing to baselines in the literature}
We compared our model to BE-DICT \citep{marquart2021predicting}. 
It is computationally intensive due to its auto-regressive sequence-to-sequence decoding and is trained as a single-task model (one model per editor). We extended and retrained BE-DICT on two randomly chosen datasets and compared its predictions with our model's results. 
\begin{table}[tb]
\centering
     \resizebox{0.9\linewidth}{!}{%
    \begin{tabular}{llllll}
    \toprule
    && \multicolumn{2}{c}{ BE-DICT}& \multicolumn{2}{c}{Ours}\\
     reference sequence & Libraries & Spearman& Pearson&Spearman&Pearson\\  
         \hline
    \multirow{2}{*}{prrotospacer}& SpRY-ABEmax & 0.801  & 0.943&0.835 &0.981\\
                                  &SpRY-ABE8e & 0.746&0.861 &0.776&0.965 \\
    \multirow{2}{*}{prrotospacer \& PAM}&SpRY-ABEmax & 0.804 &0.951 &0.870 &0.987 \\
    &SpRY-ABE8e &0.762&0.850&0.860&0.975\\
    \bottomrule
    \end{tabular}}
    \caption{\small{Performance comparison with BE-DICT}}
    \label{tab:baselines}
\end{table}
Results in Table \ref{tab:baselines} show that our model consistently outperforms BE-DICT. Furthermore, considering computational efficiency during model training 
BE-DICT takes in the order of minute per epoch, while our single-task model accomplishes the same task in the order of seconds (~15 seconds wall clock time). Notably, the multi-task learning model trained jointly on all six datasets takes ~21 seconds.

This highlights the benefits of replacing the complex sequence-to-sequence architecture in favor of a streamlined encoder-encoder structure. This choice not only improves the computational efficiency but also leads to performance enhancements. Moreover, the introduction of a two-stage model and a multi-task framework amplifies these performance gains even further. We present additional results for comparisons with other baselines in Table \ref{tab:baseline_2} in Appendix. 

\section{Conclusion}
Our work provides a detailed assessment of the modeling approaches for BE outcome prediction.
As the first machine learning-focused paper in the domain of BE outcome prediction, our work represents a stepping stone toward a systematic modeling approach to genome editing. We explored the different modeling decisions from one-stage to two-stage models, and from single-task to multi-task learning. We evaluated the different sequence representations and benchmarked our best model with one of the main models developed for base editing outcome prediction.

We believe that further work studying systematically the different modeling decisions for genome editing will help guide researchers toward more promising editing strategies that in turn will bring advancements in gene therapy.
For the future, given the current absence of standardized and systematic benchmark datasets in the field, we aim to bridge this gap by creating standard benchmark datasets, establishing baseline models, and proposing better performance metrics. This initiative will provide the machine-learning community with a solid foundation for testing a wide range of ideas.
\acks{We thank G. Schwank, K. Marquart, L. Kissling and S. Janjuha for input on the CRISPR-Cas and base editing technology and for data sharing and preprocessing. This work was supported by the URPP ‘Human Reproduction Reloaded’ and ‘University Research Priority Programs’.}
\\
\newpage
\bibliography{jmlr-main}

\begin{thebibliography}{18}
\providecommand{\natexlab}[1]{#1}
\providecommand{\url}[1]{\texttt{#1}}
\expandafter\ifx\csname urlstyle\endcsname\relax
  \providecommand{\doi}[1]{doi: #1}\else
  \providecommand{\doi}{doi: \begingroup \urlstyle{rm}\Url}\fi

\bibitem[Ameen et~al.(2021)Ameen, Ozsoz, Mubarak, Al~Turjman, and Serte]{ameen2021c}
Zubaida~Sa'id Ameen, Mehmet Ozsoz, Auwalu~Saleh Mubarak, Fadi Al~Turjman, and Sertan Serte.
\newblock C-svr crispr: Prediction of crispr/cas12 guiderna activity using deep learning models.
\newblock \emph{Alexandria Engineering Journal}, 60\penalty0 (4):\penalty0 3501--3508, 2021.

\bibitem[Arbab et~al.(2020)Arbab, Shen, Mok, Wilson, Matuszek, Cassa, and Liu]{arbab2020determinants}
Mandana Arbab, Max~W Shen, Beverly Mok, Christopher Wilson, {\.Z}aneta Matuszek, Christopher~A Cassa, and David~R Liu.
\newblock Determinants of base editing outcomes from target library analysis and machine learning.
\newblock \emph{Cell}, 182\penalty0 (2):\penalty0 463--480, 2020.

\bibitem[Ba et~al.(2016)Ba, Kiros, and Hinton]{ba2016layer}
Jimmy~Lei Ba, Jamie~Ryan Kiros, and Geoffrey~E Hinton.
\newblock Layer normalization.
\newblock \emph{arXiv preprint arXiv:1607.06450}, 2016.

\bibitem[Gaudelli et~al.(2017)Gaudelli, Komor, Rees, Packer, Badran, Bryson, and Liu]{gaudelli2017programmable}
Nicole~M Gaudelli, Alexis~C Komor, Holly~A Rees, Michael~S Packer, Ahmed~H Badran, David~I Bryson, and David~R Liu.
\newblock Programmable base editing of a• t to g• c in genomic dna without dna cleavage.
\newblock \emph{Nature}, 551\penalty0 (7681):\penalty0 464--471, 2017.

\bibitem[He et~al.(2016)He, Zhang, Ren, and Sun]{he2016deep}
Kaiming He, Xiangyu Zhang, Shaoqing Ren, and Jian Sun.
\newblock Deep residual learning for image recognition.
\newblock In \emph{Proceedings of the IEEE conference on computer vision and pattern recognition}, pages 770--778, 2016.

\bibitem[Kim et~al.(2023)Kim, Choi, Kim, Song, Seo, Min, Park, Cho, and Kim]{kim2023deep}
Nahye Kim, Sungchul Choi, Sungjae Kim, Myungjae Song, Jung~Hwa Seo, Seonwoo Min, Jinman Park, Sung-Rae Cho, and Hyongbum~Henry Kim.
\newblock Deep learning models to predict the editing efficiencies and outcomes of diverse base editors.
\newblock \emph{Nature Biotechnology}, pages 1--14, 2023.

\bibitem[Kingma and Ba(2014)]{kingma2014adam}
Diederik~P Kingma and Jimmy Ba.
\newblock Adam: A method for stochastic optimization.
\newblock \emph{arXiv preprint arXiv:1412.6980}, 2014.

\bibitem[Komor et~al.(2016)Komor, Kim, Packer, Zuris, and Liu]{komor2016programmable}
Alexis~C Komor, Yongjoo~B Kim, Michael~S Packer, John~A Zuris, and David~R Liu.
\newblock Programmable editing of a target base in genomic dna without double-stranded dna cleavage.
\newblock \emph{Nature}, 533\penalty0 (7603):\penalty0 420--424, 2016.

\bibitem[Landrum et~al.(2014)Landrum, Lee, Riley, Jang, Rubinstein, Church, and Maglott]{landrum2014clinvar}
Melissa~J Landrum, Jennifer~M Lee, George~R Riley, Wonhee Jang, Wendy~S Rubinstein, Deanna~M Church, and Donna~R Maglott.
\newblock Clinvar: public archive of relationships among sequence variation and human phenotype.
\newblock \emph{Nucleic acids research}, 42\penalty0 (D1):\penalty0 D980--D985, 2014.

\bibitem[LeCun et~al.(1995)LeCun, Bengio, et~al.]{lecun1995convolutional}
Yann LeCun, Yoshua Bengio, et~al.
\newblock Convolutional networks for images, speech, and time series.
\newblock \emph{The handbook of brain theory and neural networks}, 3361\penalty0 (10):\penalty0 1995, 1995.

\bibitem[Marquart et~al.(2021)Marquart, Allam, Janjuha, Sintsova, Villiger, Frey, Krauthammer, and Schwank]{marquart2021predicting}
Kim~F Marquart, Ahmed Allam, Sharan Janjuha, Anna Sintsova, Lukas Villiger, Nina Frey, Michael Krauthammer, and Gerald Schwank.
\newblock Predicting base editing outcomes with an attention-based deep learning algorithm trained on high-throughput target library screens.
\newblock \emph{Nature Communications}, 12\penalty0 (1):\penalty0 5114, 2021.

\bibitem[Mathis et~al.(2023)Mathis, Allam, Kissling, Marquart, Schmidheini, Solari, Bal{\'a}zs, Krauthammer, and Schwank]{mathis2023predicting}
Nicolas Mathis, Ahmed Allam, Lucas Kissling, Kim~Fabiano Marquart, Lukas Schmidheini, Cristina Solari, Zsolt Bal{\'a}zs, Michael Krauthammer, and Gerald Schwank.
\newblock Predicting prime editing efficiency and product purity by deep learning.
\newblock \emph{Nature Biotechnology}, pages 1--9, 2023.

\bibitem[Rees and Liu(2018)]{rees2018base}
Holly~A Rees and David~R Liu.
\newblock Base editing: precision chemistry on the genome and transcriptome of living cells.
\newblock \emph{Nature reviews genetics}, 19\penalty0 (12):\penalty0 770--788, 2018.

\bibitem[Song et~al.(2020)Song, Kim, Lee, Kim, Seo, Park, Choi, Jang, Shin, Min, et~al.]{song2020sequence}
Myungjae Song, Hui~Kwon Kim, Sungtae Lee, Younggwang Kim, Sang-Yeon Seo, Jinman Park, Jae~Woo Choi, Hyewon Jang, Jeong~Hong Shin, Seonwoo Min, et~al.
\newblock Sequence-specific prediction of the efficiencies of adenine and cytosine base editors.
\newblock \emph{Nature biotechnology}, 38\penalty0 (9):\penalty0 1037--1043, 2020.

\bibitem[Srivastava et~al.(2014)Srivastava, Hinton, Krizhevsky, Sutskever, and Salakhutdinov]{srivastava2014dropout}
Nitish Srivastava, Geoffrey Hinton, Alex Krizhevsky, Ilya Sutskever, and Ruslan Salakhutdinov.
\newblock Dropout: a simple way to prevent neural networks from overfitting.
\newblock \emph{The journal of machine learning research}, 15\penalty0 (1):\penalty0 1929--1958, 2014.

\bibitem[Vaswani et~al.(2017)Vaswani, Shazeer, Parmar, Uszkoreit, Jones, Gomez, Kaiser, and Polosukhin]{vaswani2017attention}
Ashish Vaswani, Noam Shazeer, Niki Parmar, Jakob Uszkoreit, Llion Jones, Aidan~N Gomez, {\L}ukasz Kaiser, and Illia Polosukhin.
\newblock Attention is all you need.
\newblock \emph{Advances in neural information processing systems}, 30, 2017.

\bibitem[Xie et~al.(2023)Xie, Liu, and Zhou]{xie2023crispr}
J~Xie, M~Liu, and L~Zhou.
\newblock Crispr-ote: Prediction of crispr on-target efficiency based on multi-dimensional feature fusion.
\newblock \emph{IRBM}, 44\penalty0 (1):\penalty0 100732, 2023.

\bibitem[Zhang et~al.(2021)Zhang, Zeng, Dai, and Dai]{zhang2021prediction}
Guishan Zhang, Tian Zeng, Zhiming Dai, and Xianhua Dai.
\newblock Prediction of crispr/cas9 single guide rna cleavage efficiency and specificity by attention-based convolutional neural networks.
\newblock \emph{Computational and structural biotechnology journal}, 19:\penalty0 1445--1457, 2021.

\end{thebibliography}

\clearpage
\section{Appendix}
\subsection{Related Work}
In recent years, the intersection of deep learning and CRISPR-Cas9 systems has witnessed substantial interest from the bioinformatics community. Researchers have explored the applications of deep learning in predicting various aspects of CRISPR-Cas9 systems, including predicting gRNA activities \citep{ameen2021c, xie2023crispr, zhang2021prediction} and editing outcomes for both base editing and prime editing scenarios \citep{mathis2023predicting}.

Among those, one notable approach is the BE-Hive proposed by \cite{arbab2020determinants}, which aims to predict base editing outcomes and efficiencies while considering sequence context, PAM compatibility, and cell-type-specific factors. The model employs a gradient boosting tree for predicting overall editing efficiency and a deep conditional autoregressive model for predicting probability of edited outcome sequences (denoted by bystander efficiency). Similarly, \cite{song2020sequence} presented DeepABE and DeepCBE, that is based on convolutional neural networks to model both overall editing efficiency and bystander efficiency of adenine and cytosine base editors. 

Recently, \citet{marquart2021predicting} proposed BE-DICT, which predicts per-base editing efficiency (i.e. editing efficiency of each target base in a sequence) and bystander base-editing efficiency using attention-based deep learning. 
In a latest comprehensive study, \cite{kim2023deep} developed DeepCas9variants and DeepBEs to predict editing efficiencies and outcomes of various BEs, taking into account different Cas9 variants. They build on and adapt the models proposed in \cite{song2020sequence} (i.e. convolutional networks) to generate predictions for a range of CRISPR-Cas9 systems.

While the surge of interest in applying machine learning to CRISPR-Cas9 systems is clear in recent literature, it's noteworthy that many of these works have a primary emphasis on designing CRISPR-Cas9 systems under various conditions and less focused on the analysis of ML models without offering a holistic and systematic analysis of model design. Given the intricate nature of CRISPR-Cas9 systems and the multitude of model paradigms adopted, deriving concrete conclusions about optimal model design strategies remains elusive. In this context, our work aims to serve as model-first work that presents the base editing outcome prediction through a modeling lens. We focus on model development and provide a systematic analysis of each component of the models, providing a structured framework for problem formulation and model design specifically tailored to the prediction of base editing outcomes.  Through this structured examination of these critical aspects, our aim is to lay the groundwork for more informed and refined approaches for using deep learning models to assist the design of base editors.


\subsection{One-stage Model}\label{sec:onestage}
In this setup, we tackle the problem by learning a function $f(\bx_{\text{ref}}, \bx_{\text{out}, i})\rightarrow \hat{y}_i$ where $i = 1, \dots, M$, and $\sum_{i=1}^M\hat{y}_i=1$, that takes as input the reference sequence and one of its corresponding outcome and learns to approximate the probability of obtaining that specific outcome. Notably, this function $f$ characterizes a categorical distribution $P(X_{\text{out}} = \bx_{\text{out}, i}|\bx_{\text{ref}})\sim Cat(M, \hat{\by})$, where $\hat{\by}$ is the vector containing probabilities for M outcomes. 
To learn the function $f$, we propose to use attention-based encoder blocks to learn the encoding of both the reference sequence and output sequence. Subsequently, we apply a prediction model on the learned encoded representation to output the probability of obtaining the outcome. The network architecture to learn $f$ is reported in figure \ref{fig:Two-step-Model} (B: proportion model). 
\subsection{Two-stage Model}\label{sec:twostage}
\subsubsection{Overall efficiency model}
We formulate the overall efficiency model as a probabilistic classification task where $f_{\mathbf{\theta}_1}$ parameterizes a binomial distribution $P(C|\bx_{\text{ref}})$ of a random variable $C\in \{\textit{edited}, \textit{not edited}\}$ with the aim to learn to output the $P(C=edited|\bx_{\text{ref}})$ for a given reference sequence. 
To learn $f_{\mathbf{\theta}_1}$, we first computed the overall editing efficiency for each reference sequence by summing all probabilities attributed to the non wild-type outcomes as given in Eq \ref{eq:efficiency}, or equivalently, $1-P(\textit{wild-type} |\bx_{ref})$. 
Then we use multiple 1D-Convolutional layers \citep{lecun1995convolutional} on the one-hot-encoded representation of $\bx_{ref}$ to learn discriminative feature embedding that is passed to the multi-layer perceptron (MLP) layer to approximate the distribution $P(C|\bx_{\text{ref}})$. 
The model architecture is presented in Figure \ref{fig:Two-step-Model} (A).
We trained $f_{\mathbf{\theta}_1}$ using KL-divergence loss that is applied on the true distribution $P(C|\bx_{\text{ref}})$  and learned distribution $\hat{P}(C|\bx_{\text{ref}})$ for each reference sequence.
\begin{align}
\resizebox{0.9\linewidth}{!}{
 $\mL_{\textit{efficiency}}(\theta_1, D) = \sum_{i=1}^N D_{kl} (P(C|\bx_{\text{ref}}^i)\|\hat{P}(C|\bx_{\text{ref}}^i))$}
\end{align}

\subsubsection{Proportion model}
This model is designed  to approximate the conditional distribution $P(X_{\text{out}}|\bx_{\text{ref}}, \textit{edited})$. To achieve this, we first remove the wild-type from each reference sequence's corresponding output $X_{\text{out}}$. Then, we normalize the probabilities of the remaining outcomes 
to ensure a valid distribution effectively converting $P(X_{\text{out}}|\bx_{\text{ref}})$ into the distribution $P(X_{\text{out}}|\bx_{\text{ref}}, \textit{edited})$. 
 The proportion model $f_{\mathbf{\theta}_2}$ is designed to learn  the parameters governing the  distribution $P(X_{\text{out}}|\bx_{\text{ref}}, \textit{edited})$. Similar to the one-stage model, $f_{\mathbf{\theta}_2}$ is provided with both the reference sequence $\bx_\text{ref}$ and its associated outcome sequence $\bx_{\text{out}, i}$. The model is then trained to estimate the likelihood $P(\bx_{\text{out}, i}\:|\:\bx_{\text{ref}}, \textit{edited})$, representing the probability of reference sequence being edited, and result in the outcome sequence $\bx_{\text{out}, i}$.

As illustrated in Figure \ref{fig:Two-step-Model} (B), $f_{\mathbf{\theta}_2}$ uses attention-based models comprised of two encoder networks, $\text{Enc}^{1}(\bx_{\text{ref}})$, $\text{Enc}^{2}(\bx_{\text{out}})$, and one output network $g$. The design of the encoder networks adapts the transformer encoder blocks architecture \citep{vaswani2017attention}, characterized by multiple layers of multi-head self-attention modules. 
The two encoder networks process the reference sequence and one of its corresponding output sequence $\bx_{\text{out}, i}$, leading to the extraction of their respective latent representations, namely $\mathbf Z_{\text{ref}}\in \mR^{T\times d}$ and $\mathbf Z_{\text{out}}\in \mR^{T\times d}$. Both vectors are then concatenated to form a unified learned representation $\mathbf Z\in\mathbb{R}^{T\times 2d}$. 
Subsequently, the output network $g$ embeds this unified representation $\mathbf Z$ to compute the probability of obtaining the output sequence given the reference sequence,  $P(\bx_{\text{out}, i}\:|\:\bx_{\text{ref}}, \textit{edited})$.

 Precisely,  the output network $g(\mathbf Z)$ takes as input the final representation $\mathbf Z\in\mathbb{R}^{T\times 2d}$  and performs an affine transformation followed by softmax operation to compute the probability of conversion  of every target base (i.e. base A or C depending on the chosen base editor) as it is shown below:
\begin{equation}
\hat{y}_{it}  = \sigma(\mathbf W\mathbf z_{it} + \mathbf b_t)
\end{equation}
where $\mathbf W\in\mR^{2\times 2d}, \mathbf b_t\in\mR^2$ and   $\sigma$ is softmax function. $\hat{y}_{it}$ represents the probability of editing occurring at the $t$-th position in the $i$-th outcome sequence. The un-normalized probability for the whole $i$-th output sequence $\bx_{\text{out}, i}$ given its reference sequence is computed by $\hat{y}_i=\prod_{t=1}^T\hat{y}_{i,t}$, which is then normalized across all the outcomes to make it valid probability distribution (Eq. \ref{eq:normalizing_prop}). Therefore, the approximated probability for obtaining $i$-th edited (non-wild type) outcome sequence is given by: 
 \begin{equation}
     \hat{P}(\bx_{\text{out}, i}\:|\:\bx_{\text{ref}}, \textit{edited}) =\frac{\hat{y}_i}{\sum_{i=1}^M\hat{y}_i}
    \label{eq:normalizing_prop}
 \end{equation}

\paragraph{Objective Function}
We used the Kullback–Leibler (KL) divergence on the model’s estimated distribution over all outcome sequences for a given reference sequence $\bx_{\text{ref}}^i$ and the actual distribution:
\begin{flalign}
    &D_{\text{KL}}^i(P(X_{\text{out}}|\bx_{\text{ref}}^i, \textit{edited})||\hat{P}(X_{\text{out}}|\bx_{\text{ref}}^i, \textit{edited}))\\
    &= \sum_{j=1}^{M_i} P(\bx_{\text{out},j}|\bx_{\text{ref}}^i, \textit{edited})\log \frac{P(\bx_{\text{out},j}|\bx_{\text{ref}}^i, \textit{edited})}{\hat{P}(\bx_{\text{out},j}|\bx_{\text{ref}}^i, \textit{edited})} \nonumber
\end{flalign}
Lastly, the objective function for the whole training set is defined by the average loss across all the reference sequences as follows:
\begin{align}
    &\mL_{\text{proportion}}(\mathbf{\theta_2}; D) =\\
    &\sum_{i = 1}^N D_{KL}^i (P(X_{\text{out}}\:|\:\bx_{\text{ref}}^i, \textit{edited})||\hat{P}(X_{\text{out}}\:|\:\bx_{\text{ref}}^i, \textit{edited}) \nonumber
\end{align}


\subsection{Model architecture}\label{sec:network_architecture}
\subsubsection{Single-task learning}\label{single-task-learning}
In this paper, we refer to single-task learning as a setting where we train one separate model for each of the libraries. The terminology is used to contrast with multi-task learning where we train one unified model for all the editors/datasets. For the single-task learning, we used the two-stage model ( Figure \ref{fig:Two-step-Model}) with protospacer and PAM as the reference sequence representation.  In this section, we provide an in-depth introduction to the two-stage model architecture, which is comprised of two distinct sub-models: the overall efficiency model and the proportion model.

\paragraph{Overall Efficiency Model}
The Overall Efficiency Model concentrates exclusively on the target sequence, overlooking the specific edit outcomes. Its main objective is to predict the probability of the target sequence undergoing modification, regardless of the nature of the edits. Hence, the model exclusively processes the input target sequence, which in our scenario is the concatenation of the protospacer and PAM, yielding the probability of the target sequence undergoing modification (yielding non-wild type outcomes). To achieve this, we propose to use a Convolutional Neural Network (CNN) on the one-hot encoding of the target sequence. More specifically, we use three layers of 1D-CNN (kernel size: 2, stride: 2) with filter sizes of 32, 64, and 128, respectively. Following the CNN layers, we apply a feed-forward network with ReLU activation, featuring a hidden layer dimension of 64. The output of this network is a two-dimensional vector, which is subsequently transformed into probabilities through the use of the Softmax function.

\paragraph{Proportion Model}
Different from the absolute efficiency model, the proportion model focuses on predicting the probability of different types of edited outcomes for the target sequence. Therefore, it takes both the target sequence as well as one of its corresponding outcome sequences and outputs the probability of observing such an outcome. To implement this model, we use two encoder networks and one prediction network. The architectures of the two encoding networks, one for the target sequence and the other for the outcome sequence are identical, as illustrated in Figure \ref{fig:Two-step-Model}. Consequently, we will only describe in detail one of these networks here for clarity.
The target/reference sequence encoder network comprises two essential components: an embedding block and an encoder block.

\paragraph{Embedding Layer}
The embedding block embeds both the nucleotides and their corresponding position (in the protospacer) from the one-hot encoded representation to a dense vector representation. Given a protospacer sequence extended with its corresponding PAM site: $\bx_{\text{ref}} = [x_1, x_2, \dots,x_T]\in \mR^{T}$, $x_t$ represents the nucleotide at position $t$. In our case, T=24. We use $\mathbf O=[\mathbf o_1, \mathbf o_2, \dots, \mathbf o_T]\in \mR^{K\times T}$ as its one-hot encoded representation.  Here $K = 4$ as we have only four distinct nucleotides.  

An embedding matrix $\mathbf W_e$ is used to map each $\mathbf o_t\in \mR^k$ to a fixed-length vector representation:
\begin{equation}
    \mathbf e_t = \mathbf W_e\mathbf o_t
\end{equation}
where $\mathbf W_e\in \mR^{d_2\times K}$, $\mathbf e_t \in \mR^{d_e}$,  and $d_e$ is the embedding dimension we chose. 

Similarly, each nucleotide's position in the sequence $\bx_{\text{ref}}$ is represented by one-hot encoding with dictionary size $T$. En embedding matrix $\mathbf W_{p'}\in\mR^{d_e\times T}$ is applied to project the  $\mathbf p_4$ to a dense vector representation:
\begin{equation}
     \mathbf p'_t = \mathbf W_{p'}\mathbf p_t
\end{equation}
 where $\mathbf W_{p'}\in \mR^{d_2\times T}$, $\mathbf p_t \in \mR^{d_e}$. 
 Both embeddings $\mathbf e_t$ and $\mathbf p'_t$ are summed to get a unified representation for every element $x_t$ in the reference sequence $\bx_{\text{ref}}$. 
 \begin{equation}
     \mathbf u_t = \mathbf e_t + \mathbf p'_t \:\: \:\: \forall \:\:  t= 1, 2, \dots T
 \end{equation}
 This results in the embedded representation $\mathbf U=[\mathbf u_1, \mathbf u_2, \dots, \mathbf u_T]$ of the reference sequence. 

\paragraph{Encoder Block}
To learn a good representation that takes into account the relationships between the nucleotides in the reference sequence, we use a multi-head self-attention to encode the embedded representation. Multi-head Attention is a module that employs multiple single-head self-attention in parallel (i.e. simultaneously) to process each input vector $\mathbf u_t$. The outputs from every single-head layer are then concatenated
and transformed by an affine transformation to generate a fixed-length vector.

The single-head self-attention approach \citep{vaswani2017attention} learns three different linear projections of the input vector using  three separate  matrices: (1)
a queries matrix $\mathbf W_{query}$, (2) keys matrix $\mathbf W_{key}$, and (3) values matrix $\mathbf W_{value}$. Each input $\mathbf u_t$ in $\mathbf U$ is mapped
using these matrices to compute three new vectors:
\begin{align}
   & \mathbf q_t = \mathbf W_{query}\mathbf u_t\\
    &\mathbf k_4 = \mathbf W_{key}\mathbf u_t\\
    &\mathbf v_t = \mathbf W_{value}\mathbf u_t\\
\end{align}
where $\mathbf W_{query}$, $\mathbf W_{key}$, $\mathbf W_{value}\in \mR^{d\times d_e}$, $\mathbf q_t, \mathbf k_t, \mathbf v_t \in \mR^{d}$ are query, key and value vectors, and $d$ is the dimension of the those projected vectors. In the second step, attention scores are computed using the pairwise similarity between the query and key vectors for each position $t$ in the sequence. The similarity is defined by
first computing a scaled dot product between the pairwise vectors and then normalizing it using the softmax function. At each position $t$, we compute attention scores
$\alpha_{tl}$ representing the similarity between $t$-th query $\mathbf q_t$ and $l$-th key $\mathbf k_l$.
\begin{equation}
    score(\mathbf q_t, \mathbf k_l) = \frac{\mathbf q_t^T\mathbf k_l}{\sqrt{d}}
\end{equation}
\begin{equation}
    \alpha_{tl} = \frac{\exp(score(\mathbf q_t, \mathbf k_l))}{\sum_{l=1}^T \exp(score(\mathbf q_t, \mathbf k_l))}
\end{equation}
Then a weighted sum of value vector $\mathbf v_l$ using attention $\alpha_{tl}, \:\: \forall \: l\in \{1,2,\dots, T\}$ is performed to generate a new vector representation $\mathbf e_t\in \mR^d$ at position $t$. 

\begin{equation}
    \mathbf e_t = \sum_{l=1}^T \alpha_{tl}\mathbf v_l
\end{equation}
This process is applied to every position in the original embedding of the sequence, $\mathbf U$, to obtain a sequence of vectors $\mathbf E = [\mathbf e_1, \mathbf e_2, \dots, \mathbf e_T]$.

In a multi-head setting with H number of heads, the queries, keys, and values matrices will be indexed by superscript $h$ (i.e. $\mathbf W_{query}^h$, $\mathbf W_{key}^h$, $\mathbf W_{value}^h\in \mR^{d\times d_e}$) and applied separately to generate a new vector representation $\mathbf e_t^h$ for every singe-head self-attention layer. The output from each single-head attention layer is contenated into one vector $\mathbf e_t^{concat}=concat(\mathbf e_t^1, \mathbf e_t^2, \dots, \mathbf e_t^H)$ where $\mathbf e_t^{concat}\in \mR^{dH}$. Then it goes through an affine transformation using $\mathbf W\in \mR^{d\times dH}$ and $\mathbf b\in \mR^d$ to generate the encoded representation $\hat{\mathbf{Z}} = [\hat{\bz}_1, \hat{\bz}_2, \dots,\hat{\bz}_T]$ of the reference sequence
\begin{equation}
    \hat{\bz}_t = \mathbf W\mathbf e_t^{concat} + \mathbf b
\end{equation}

To improve the gradient flow in layers during
training, we also use residual connections / skip-connections \citep{he2016deep}. This is done by summing both the newly computed output of the current layer with the output from the
previous layer. In our setting, a first residual connection sums the output of the self-attention layer $\hat{\bz}_t$ and the
output of embedding block $\mathbf u_t$ for each position $t$ in the sequence.

We also deploy layer normalization \citep{ba2016layer} after the self-attention layer 
with the goal of ameliorating the "covariate-shift" problem by re-standardizing the computed vector representations
(i.e. using the mean and variance across the features/embedding dimension $d$). Given a computed
vector $\hat{\bz}_t$, the LayerNorm function will standardize the input vector using the mean and variance
along the dimension of the feature $d$ and apply scaling 
 and shifting steps.

Eventually, this learned representation goes through a feedforward network with one hidden layer and ReLu activation function. Subsequently, a layer normalization is applied to the output of this feed-forward network to obtain the learned representation $\bz_t\in \mR^d$. Eventually, the encoder block transformed the embedded vector $U$ to the learned representation $\mathbf Z=[\bz_1, \bz_2, \dots, \bz_T]\in \mR^{T\times d}$ that incorporates the contextual information/relationships between features through attention.

The above process describes one encoding block, we stack N such blocks to construct our encoder network. As it is presented in Figure \ref{fig:Two-step-Model}, in the proportion model, we apply two encoder networks with the same network architecture on the reference sequence and corresponding outcome sequence. This yields two encoded representations which we denote by $\mathbf Z^{ref}\in \mR^{T\times d} $ and $\mathbf Z^{out}\in \mR^{T\times d} $ respectively. We then concatenated them in the feature dimension to generate one common representation $\mathbf Z= [\bz_1, \bz_2, \dots, \bz_T]\in \mR^{T\times 2d}$. 

\paragraph{Output network} 
The output network consists of an affine transformation and a nonlinear function to transform the output to probability:

\begin{equation}
    \by_{out,t} = \mathbf W_{out}\bz_t
\end{equation}
where $\mathbf W_{out}\in \mR^{2d\times 2}$,$\by_{out,t}\in \mR^2$. We then apply a softmax function on $\by_{out,t}$ and transform it to a probability that represents the probability of the nucleotide at position $t$ getting edited. Note that, we use the same length of input and outcome sequence. However, our input sequence also includes PAM information, and editing only happens in the protospacer. Moreover, due to the nature of the Base editor, only specific nucleotides, in our case Adenine (A) gets edited while the other nucleotides are not affected. Therefore, we use masking technique to only consider the positions that are possible to be edited and mask out other positions. Therefore, the PAM information (or the contextual information such as left/right overhangs is participating by affecting the embedding of the nucleotides in the protospacer but is not considered in the loss as they are not changed/edited.

After tuning the parameters, for the Proportion Model, we have chosen an embedding dimension of 124 for the embedding layer. Our model consists of 12 encoding blocks, with each block featuring 8 multi-head attention mechanisms. For the output network, we employ a single linear layer that maps a 248-dimensional vector to a two-dimensional output vector. Subsequently, we transform this output into probabilities using the softmax function.

\subsection{Multi-task Learning }
We extended the two-stage model for accommodating various base editors through the implementation of a multi-task learning framework, eliminating the need for training individual models per base editor.

To achieve this, we first augment the datasets from different libraries with corresponding editor labels and combine all the libraries to create a consolidated dataset. Our objective is to establish a shared architecture comprising common layers applicable across all libraries, along with dedicated sub-networks tailored to each specific library. 

As illustrated in Figure \ref{fig:multi_task}, we begin by extending the overall efficiency model by incorporating the first two layers of a 1D-CNN as the universally shared layers for all libraries. Subsequently, this shared learned representation traverses a sub-network comprising two layers of 1D-CNN and two layers of MLP with ReLU activation functions, uniquely customized for each library. Given that, as we have datasets from six editors, we use one common network (consisting of two layers of CNN)  and six distinct sub-networks, which have identical structures. 

For the proportion model, the influence of various libraries/Base Editors is observable through the variations in the outcome set corresponding to the various base editors when applied to the same target sequence. Therefore, to extend the proportional model for various base editors, we maintain a shared encoder network for the reference sequence across all libraries while constructing six distinct encoder networks to encode the outcome sequences from the six different libraries. This approach allows us to establish a consistent representation for the reference sequences across all libraries, while simultaneously accommodating the distinctions among the libraries by employing separate encoder networks to encode the outcome sequences for each one. Consequently, our multi-task learning proportional model comprises one shared reference sequence encoder network, six individual outcome sequence networks, and six corresponding output networks.
\subsection{Optimization}
For optimization, we use Adam optimizer \citep{kingma2014adam} with a learning rate scheduler. We initialize the base learning rate at $3e^{-4}$ and set the maximum learning rate to five times the base rate.  Additionally, we incorporate dropout \citep{srivastava2014dropout} probability of 0.2. The regularizer parameter $\lambda $ is set to $1e^{-4}$. 

It's worth noting that while it's possible to train both the overall efficiency model and the proportion Model simultaneously. However, 
Training them together means, we apply the loss on the learned final probability,  $\hat{P}(X_{out}|\bx_{ref})$ (Eq. \ref{model:full}), with the true probability which represents the probability of all outcomes. This means that for the edited (non-wild outcome) outcomes, the problem of true probability being very low compared to the wild-type outcomes still exists. Breaking down this final probability into a product of two conditional probabilities was introduced to mitigate such problem as $P(X_{out}|\bx_{ref}, edited)$ focuses only on the non-wild type outcomes. Therefore, training them together could result in an outcome where the model predominantly focuses on predicting easily predictable outcomes (as wild type) while neglecting those with lower probabilities as we hypothesize in the One-stage model. Moreover, the two models do not share any common layer except the outcome of the two networks gets multiplied to generate the final probability distribution. Therefore, there is no real requirement by model design to train them together. To avoid falling back to the single-step model, we train the Absolute Efficiency Model and the Proportion Model separately. This approach explicitly matches the probabilities $P(C|\bx_{\text{ref}})$ and $P(X_{out}|\bx_{\text{ref}}, \text{edited})$, preventing the model from overlooking low-probability outcomes.

In terms of training specifics, we set the mini-batch size to 100 and the maximum number of epochs to 300 for the Absolute Efficiency Model. For the Proportion Model, we choose a mini-batch size of 400 and an epoch count of 150. In both models, we use Spearman correlation on the validation set as our performance metric to monitor and select the best-performing model.

\subsection{Performance measures} 
Considering the distinctive nature of our data generation process, we chose to employ Pearson and Spearman correlations as our performance metrics, measuring the alignment between actual and predicted probability scores. Owing to inherent variability during screening, repeated experiments under identical conditions yield slightly divergent outcomes. For the biologist, the emphasis is on correlating these similar results, rendering metrics such as mean square or mean absolute error less pertinent. Our primary concern isn't exact prediction precision, but rather the level of correlation achieved between predictions and actual data.

\subsection{CRISPR related terminology}\label{crispr}
The CRISPR-Cas9 system is a revolutionary gene-editing technology that allows scientists to precisely modify DNA within living organisms. Initially, it was described as an adaptive immune system of bacteria and archea to eliminate invading foreign DNA and/or RNA. The system uses unique sequences of RNA called single guide RNA (sgRNA) that are recognized and bound by Cas9, an enzyme with a nuclease domain. The Cas9 protein carries out the initial steps of recognition and binding by scanning genomic DNA to locate a particular sequence called a PAM (protospacer adjacent motif). Upon PAM recognition, the part of the sgRNA (termed spacer) complementary to the target DNA (termed protospacer) opens the DNA double helix and binds to the target site. This leads to a conformation change within the Cas9, bringing its nuclease domain in close proximity to the target DNA and thus initiating DNA double-strand cleavage.
After introducing a DNA double-strand break, the cell's repair mechanism is triggered, which can lead to various outcomes. Researchers can exploit this repair process to either introduce specific changes in the DNA sequence by providing a modified DNA template or to disrupt a target gene due to the imperfect repair of the DNA by the cells, thus introducing insertions or deletions, which can lead to a frame shift mutations.

Of note, Streptococcus pyogenes Cas9 (SpCas9) exclusively operates on ``NGG'' (``N'', any base) PAM sequence. Recent efforts in protein design have resulted in laboratory-generated SpCas9 variants, such as SpG or SpRY, which are able to recognize different PAM motifs. Therefore, the editors used in our high-throughput screening are configured to operate with PAM sequences comprising four nucleotides.


\subsubsection{Base Editor (BE)}\label{apend:base_editor}
Base editing \citep{komor2016programmable, gaudelli2017programmable, rees2018base} is a second-generation genome editing approach that uses components from CRISPR systems together with other enzymes to directly install point mutations into genomic DNA without making double-stranded DNA breaks (DSBs). BEs comprise a Cas protein with a catalytically impaired nuclease domain fused to a nucleobase deaminase. Similar to the Cas9 nuclease, BEs are directed to the target DNA by the programmable sgRNA and are able to directly convert substrate bases in a specific 'editing window' within the protospacer.



\subsection{Description of Each Base Editor Used in the Experiment}\label{appendix:BE}
There are two main factors that are crucial for the efficiency of the base editor: binding affinities and deaminase activity. 
Binding affinities dictate how effectively an editor is able to identify and interact with specific target sites on the reference sequence. Editors with higher binding affinities tend to exhibit increased accuracy in achieving the desired base modification. Deaminase activity is defined by the type of deaminase used. Additionally, the deaminase also defines editing window size, which refers to the span of nucleotides that an editor can modify around its target site. Editors with larger editing windows can potentially influence a broader range of nucleotides, resulting in increased flexibility in terms of target selection and outcomes.

Here we describe six editors that we used in our screening experiment. We used three different Cas9 orthologs, which show different binding affinities to different PAMs: SpCas9, SpG, and SpRY. SpCas9 recognizes only a few PAMs but shows a very high affinity for those PAMs. SpRY shows the broadest PAM recognition, however a lower affinity for all of them. SpG recognizes more PAM than SpCas9, however less than SpRY and shows lower affinity than SpCas9 but higher as SpRY. For deaminases we used two evolved adenine deaminases named ABEmax and ABE8e, where ABE8e has higher deaminase activity and a bigger editing window size. The combination of the three Cas9 orthologs and the two adenine deaminases leads to a total of six adenine base editors.


\subsection{Data Statistics}
\begin{table}[tb]
    \centering
    \resizebox{\linewidth}{!}{%
    \begin{tabular}{llllll}
    \toprule
         Editor & \#ins&\#refseq& \#outcome&mean 
         &std  \\
         \hline
         SpRY-ABE8e&110141&11291 &9.7&0.102&0.211\\
        
         SpCas9-ABE8e&43054&11337 &4.6&0.217&0.323\\
         
         SpG-ABE8e&80873&11307&7.1&0.139&0.263\\
        SpRY-ABEmax&70851&11347&6.2&0.159&0.301\\
        SpCas9-ABEmax&39606&11302&3.5&0.285&0.417\\
        SpG-ABEmax&70851&11347&6.2&0.159&0.301\\
\bottomrule
    \end{tabular}}
    \caption{\small{Data statistics: ``\#ins" refers to the number of reference and output sequence pairs, ``\#refseq" denotes the number of distinct reference sequences, ``\#outcome" denotes the average number of outcomes per reference sequence, the mean and std refers to the mean and standard deviation of the probability across all the outcomes.}}
    \label{tab:data}
\end{table}

\subsection{Experiment results}
\begin{table*}
    \centering
     \resizebox{0.8\textwidth}{!}{%
    \begin{tabular}{lllllll}
    \toprule
    & \multicolumn{2}{c}{Protspacer}& \multicolumn{2}{c}{Protospacer  \& PAM}&\multicolumn{2}{c}{Protspacer \& PAM \& Overhangs}\\
      Libraries & Spearman& Pearson&Spearman&Pearson&Spearman& Pearson\\
         \hline
         SpRY-ABEmax    &$0.835 \pm 0.007$&$0.981 \pm 0.001$& $0.854 \pm 0.006$&$0.983 \pm 0.001$&$0.854 \pm 0.003$&$0.983 \pm 0.002$ \\
         SpCas9-ABEmax &$0.786 \pm 0.003$&$0.978 \pm 0.002$&$0.881 \pm 0.001$&$0.989 \pm 0.0005$&$0.891 \pm 0.002$&$0.989 \pm 0.001$\\
         SpG-ABEmax&$0.841 \pm 0.002$&$0.985 \pm 0.0007$&$0.866 \pm 0.004$&$0.989 \pm 0.0003$&$0.878 \pm 0.008$&$0.991 \pm 0.0009$\\
        SpRY-ABE8e &$0.776 \pm 0.019$&$0.965 \pm 0.001$&$0.779 \pm 0.0036$&$0.968 \pm 0.002$&
$0.803 \pm 0.008$&$0.967 \pm 0.0003$\\
        SpCas9-ABE8e&$0.762 \pm 0.007$&$0.883 \pm 0.005$&$0.857 \pm 0.007$&$0.945 \pm 0.0006$&$0.862 \pm 0.003$&$0.945 \pm 0.003$\\
        SpG-ABE8e&$0.803 \pm 0.005$&$0.963 \pm 0.002$&$0.820 \pm 0.005$&$0.974 \pm 0.0009$&$0.819 \pm 0.006$&$0.9771 \pm 0.0008$ \\
    \bottomrule
    \end{tabular}}
    \caption{\small{Pearson and Spearman correlation using one-stage Model across the three different reference sequence representations. In our experiment,  we chose 5 neighboring nucleotides for both sides to represent the overhangs. }}
    \label{table:reference_seq}
\end{table*}

\subsubsection{Reference sequence representation}\label{sec:input_rep}
Existing models have explored different factors that could affect the base editor's efficiency, which we categorize into three scenarios: 1) the protospacer, 2) the protospacer along with its PAM, and 3) an extended range including left overhangs, protospacer, PAM, and right overhangs. We investigate all three scenarios with the one-stage model to identify the best features to represent the reference sequence. We observe that (See Table \ref{table:reference_seq} in appendix), incorporating PAM information significantly enhances performance, whereas the inclusion of overhangs demonstrates minimal impact. Besides, adding overhangs increases the computational complexity drastically. Consequently, we opt to employ protospacer and PAM information to represent reference sequences in all the subsequent model results presented below.

\subsection{Comparing with the other baselines}\label{sec:baseline_comparision}
To assess our model's performance against other state-of-the-art models, we conducted evaluations using the test sets provided by these models. Table \ref{tab:baseline_2} displays our findings, which include three most recent models: BE-HIVE \citep{arbab2020determinants}, DeepABE \citep{song2020sequence}, and BEDICT \citep{marquart2021predicting}, along with their respective test sets labeled as A. et al., S. et al., and M. et al.

\begin{table}[tb]
    \centering
    \resizebox{\linewidth}{!}{%
    \begin{tabular}{lllllll}
    \toprule
    &\multicolumn{3}{c}{ All Outocmes}& \multicolumn{3}{c}{Non wild-types}\\
         Datasets& A.et all&S. et al &M. et al& A.et all&S. et al &M. et al  \\
         \hline
     BEDICT & 0.96& 0.94&0.86&0.81& 0.90&0.82\\
     DeepABE&0.86&0.93&0.8&0.86&0.96&0.84\\
     BE-HIVE&0.71&0.88&0.74&0.92&0.93&0.81\\
     Our model & 0.972&0.974&0.972& 0.939&0.945&0.953\\
     \hline
    \end{tabular}}
    \caption{\small{Model performance on the test set from the different published studies. Columns represent test sets, rows represent models used}}
    \label{tab:baseline_2}

\end{table}
The idea is to take the published trained model and evaluate their performance on those various test sets. For the three baseline models, we refer to the results reported in the BEDICT paper. As for our model, to ensure fairness in comparison, we used our single-step model trained on SpG-ABEmax libraries\footnote{other ABEmax libraries also yield similar results} since most baselines, except DeepABE, do not incorporate the PAM as input. The results correspond to two scenarios: 1) considering all possible outcomes, and 2) only considering non-wild type outcomes. The results for the non-wild type outcomes correspond to the model prediction where we only consider non-wild outcomes.  In the case of non-wild-type outcome prediction, we mention that other models were trained exclusively on non-wild outcomes, with outcomes per sequence being renormalized. Our one-stage model, however, was trained on data encompassing all outcomes, so we report non-wild-type results with outcomes renormalized for a fair comparison.



\subsubsection{Multi-task learning}
 To extend our two-stage model in the setting of multi-task learning (\ref{sec:Multi_task_learning}), we explored two distinct methodologies for tackling multi-task learning. The first involves a direct conversion of the distribution into a conditional form, conditioned upon the editor label. The second applies a structural transformation of the network enabling the model to have both shared and distinct layers across various libraries. We refer to the first as a conditional model and the second as multi-task learning. 
 In order to identify a suitable approach, we exclusively assessed both methodologies using the absolute efficiency model, leveraging its inherent simplicity. This choice stems from the rationale that if the conditioning factor is overlooked within this inherently simpler context, its impact is likely to be minimal when applied to the proportion model, which is considerably more intricate. As illustrated in Table \ref{tab:conditional_model}, the multi-task setup on the absolute efficiency model has a substantial advantage over the model that uses the editor label as a conditioning factor.

\begin{table}[tb]
    \centering
      \resizebox{\linewidth}{!}{%
     \begin{tabular}{lllll}
     \toprule
     & \multicolumn{2}{c}{ Conditional model}& \multicolumn{2}{c}{Multi task learning}\\
       Libraries & Spearman& Pearson&Spearman&Pearson\\  
          \hline
     SpRY-ABEmax & $0.677 \pm 0.004$ & $0.629 \pm 0.003$ & $0.797 \pm 0.007$ & $0.783 \pm 0.012$ \\
    SpCas9-ABEmax & $0.811 \pm 0.009$ & $0.759 \pm 0.002$ & $0.834 \pm 0.011$ & $0.901 \pm 0.002$ \\
     SpG-ABEmax & $0.811 \pm 0.009$ & $0.748 \pm 0.006$ & $0.853 \pm 0.009$ & $0.835 \pm 0.017$ \\
     SpRY-ABE8e & $0.548 \pm 0.012$ & $0.537 \pm 0.018$ & $0.578 \pm 0.031$ & $0.695 \pm 0.034$ \\
     SpCas9-ABE8e & $0.751 \pm 0.010$ & $0.723 \pm 0.016$ & $0.866 \pm 0.031$ & $0.862 \pm 0.014$ \\
     SpG-ABE8e & $0.788 \pm 0.005$ & $0.760 \pm 0.011$ & $0.788 \pm 0.002$ & $0.824 \pm 0.004$ \\
     \bottomrule
     \end{tabular}}
     \caption{\small{Performance comparison of overall efficiency model on two different settings: conditional model $P(C|\bx_{\text{ref}}, B_i)$ and multi-task learning approach }}
\label{tab:conditional_model}
\end{table}
\subsection{Scatter Plots}
To gain deeper insights into the model's performance, we provide scatter plots showcasing the actual and predicted probability values derived from the multi-task model on the SpRY-ABEmax library test set. It's important to note that the library selection is entirely random, and this particular library is not cherry-picked; similar results are observed across all other libraries as well. 

\begin{figure}[tb]
    \centering
    \includegraphics[scale=0.5]{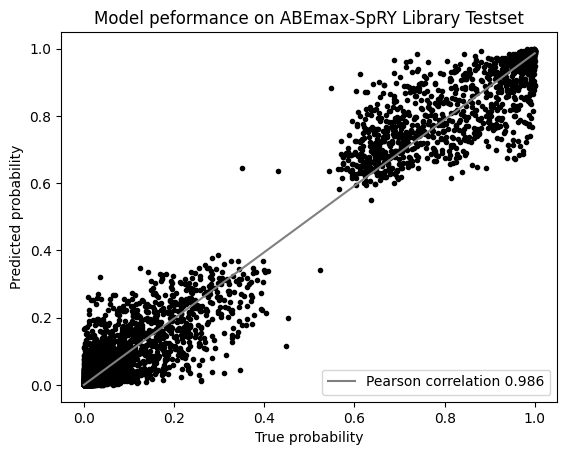}
    \caption{Performance of the Multi-Task Model Across All Possible Outcomes, Including Both Wild-Type and Non-Wild-Type Variants}
    \label{fig:scater1}
\end{figure}
\begin{figure}[tb]
    \centering
    \includegraphics[scale=0.5]{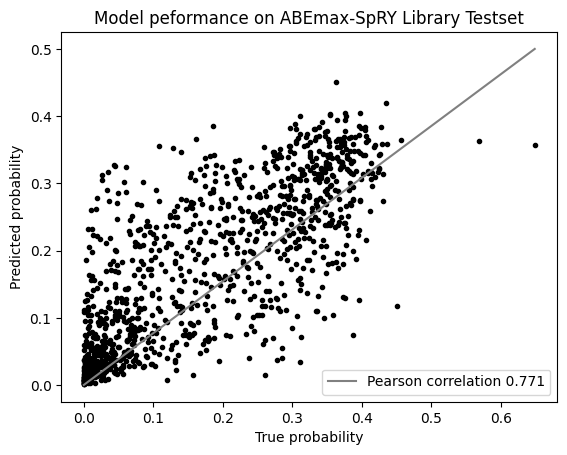}
    \caption{Performance of the Multi-Task Model Across  Wild Type (this corresponding to the model absolute efficiency model performance)}
    \label{fig:scater2}
\end{figure}
\begin{figure}
    \centering
    \includegraphics[scale=0.5]{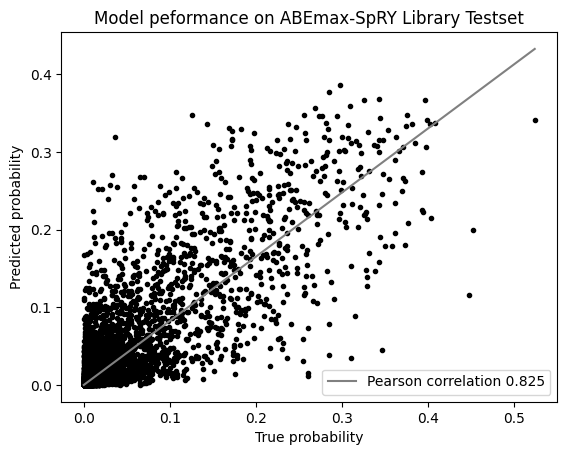}
    \caption{Performance of the Multi-Task Model Across Non-Wild Type}
    \label{fig:scater3}
\end{figure}
\begin{figure}[tb]
    \centering
    \includegraphics[scale=0.5]{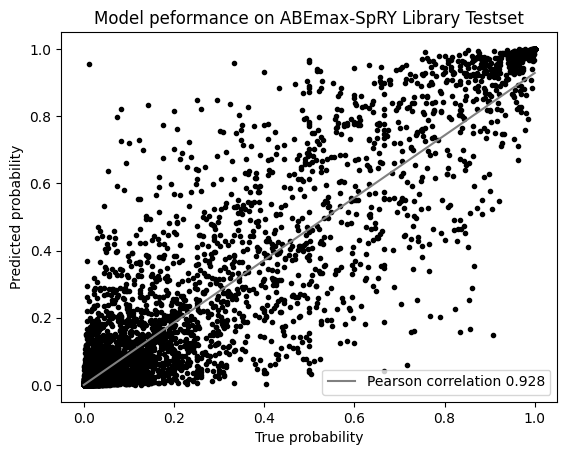}
    \caption{Performance of the Multi-Task Model Across Non-Wild Type (presented in the randomized space, i.e., $P(X_{out}|\bx_{\text{ref}, edited}$, this corresponds to the proportional model performance)}
    \label{fig:scater4}
\end{figure}
\newpage

\end{document}